\documentclass[a4paper,11pt]{article}
\usepackage{pos}
\newcommand{\MG} {\textsc{mg5}\_a\textsc{mc}}

\newcommand*{\ttbar}{$\mathrm{t}\overline{\mathrm{t}}$}

\newcommand*{\alpS}{$\alpha_S$}

\newcommand*{\alpFSR}{$\alpha_S^{FSR}$}
\newcommand*{\SHERPA} {{\textsc{sherpa}}}
\newcommand*{\PYTHIA} {{\textsc{pythia8}}}
\newcommand*{\POWHEG} {{\textsc{powheg}}}
\newcommand*{\HERWIG} {{\textsc{herwig7}}}

\newcommand*{\EPOS}{{\textsc{epos}}}
\newcommand*{\QGSJET}{{\textsc{qgsjet-ii}}}

\newcommand*{\pt}{\ensuremath{p_{\mathrm{T}}}}
\newcommand{\CP}{\textsc{cuetp8m2t4}}

\newcommand*{\mt}{$m_{\rm t}$}

\title{Modelling the Data at the LHC}

\ShortTitle{Modelling the Data at the LHC}

\author*[a,b]{Efe Yazgan}

\affiliation[a]{on behalf of the ATLAS and CMS Collaborations}

\affiliation[b]{National Taiwan University,\\
  Department of Physics, Laboratory of High Energy Physics, 10617 Taipei, Taiwan}
  
\emailAdd{efe.yazgan@cern.ch}

\abstract{
Measurements at hadron colliders rely on large scale quantum chromodynamics (QCD) Monte Carlo (MC) production for interpretation of the data. 
MC simulations allow testing Standard Model (SM) with more accurate and precise calculations to understand perturbative QCD as well as electroweak effects, and extrapolations of the irreducible backgrounds to signal phase-space regions for new physics searches or for the measurements of rare SM processes.
In the MC codes, there are many pieces, approximations, and parameters and settings to compare to the data and tune. 
Precise experimental measurements at the LHC require similar level of precision in theoretical calculations. 
Cross sections measured at the LHC both by ATLAS and CMS experiments cover more than 14 orders of magnitude. 
So far, SM cross section predictions are found to be in very good agreement with the data. 
These cross sections are measured at different pp collision energies and compared to prediction up to next-to-next-to-leading order (NNLO) for many processes, and recently up to N$^3$LO for some of them. In this note, a few measurements relevant to data modelling are discussed.  
}

\FullConference{%
  *** Particles and Nuclei International Conference - PANIC2021 ***\\
  *** 5 - 10 September, 2021 ***\\
  *** Online ***
}

\begin{document}
\maketitle

\section{Total Cross Sections}
\begin{figure}[ht]
    \centering
    \includegraphics[width=7.5cm]{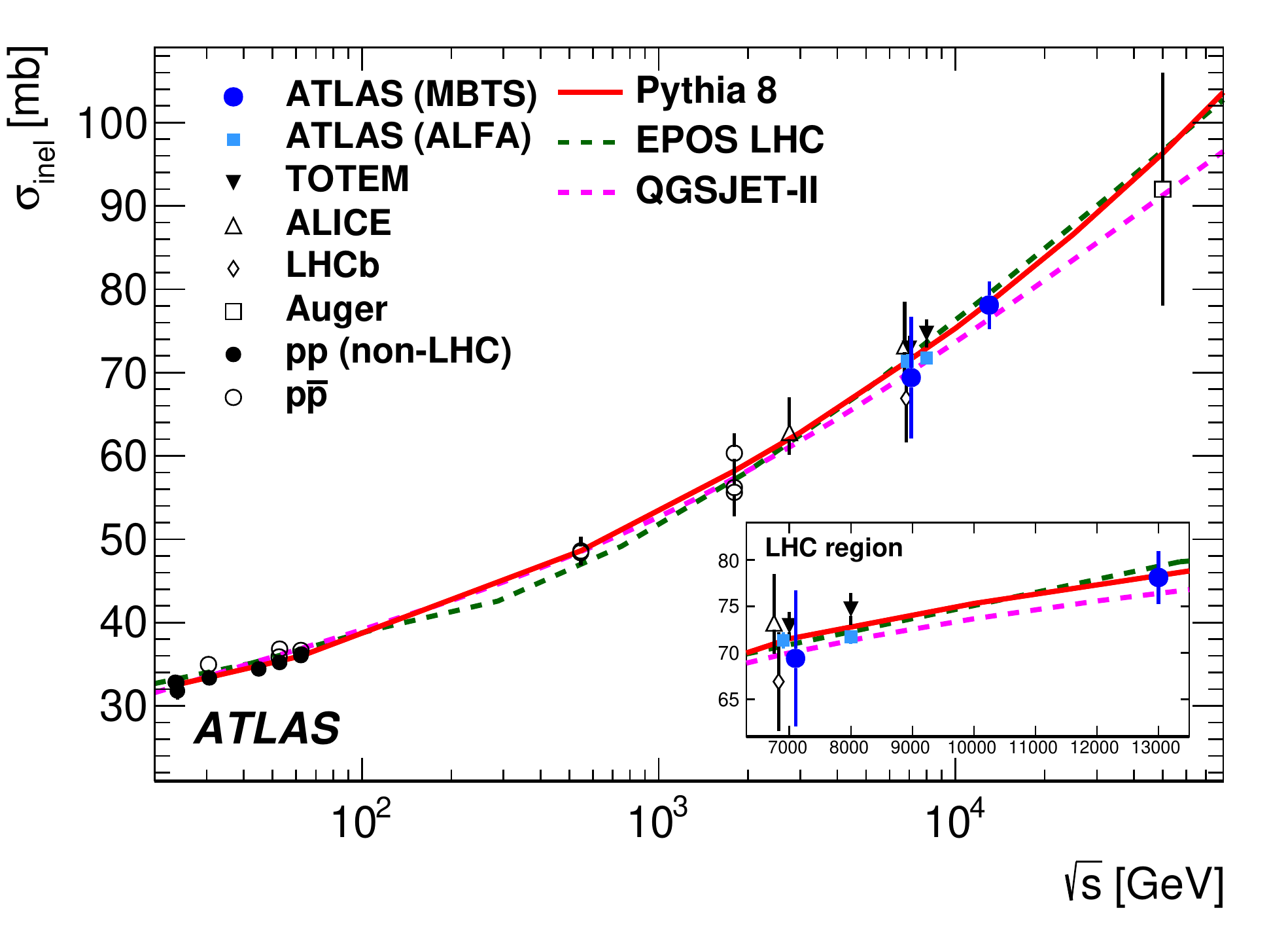}
    \caption{The inelastic proton-proton cross section as a function of $\sqrt{s}$. Measurements from other hadron collider experiments and the Pierre Auger observatory are also shown. See Ref.~\cite{ATLAS:2016ygv} for the corresponding references.}
    \label{fig:totalcrosssection}
\end{figure}

The increase of total proton-proton (pp) cross sections with center-of-mass energy ($\sqrt{s}$) is first observed at the CERN ISR in 1973~\cite{amaldi73}. The total pp cross sections can not be calculated using perturbative quantum chromodynamics (QCD), however,
unitarity, analyticity, and factorization arguments suggest that the total hadronic cross sections should rise slower than $ln^2(s)$~\cite{froissart61}. 
ATLAS~\cite{ref:ATLAS} measured the inelastic pp cross section at $\sqrt{s}=13$ TeV by selecting events with rings of plastic scintillators in the forward region (2.07$<$$|\eta|$$<$3.86)~\cite{ATLAS:2016ygv} in the fiducial phase space defined by $M^2_x/s$ $>$10$^{-6}$ to be 68.1$\pm$1.4 mb. Here, $M^2_x$ is the larger of the invariant mass any of the two final-state hadrons separated by the largest rapidity gap in the event. Using the $\sqrt{s}=7$ TeV measurements, this  measurement is extrapolated to obtain the total inelastic cross section of 78.1$\pm$2.9 mb. The measurements are in good agreement with \PYTHIA~\cite{pythia8} and \EPOS~\cite{Pierog:2013ria} in the LHC region and are consistent with the inelastic cross section increasing logarithmically with $\sqrt{s}$ up to $\sim$60 TeV also including the cosmic ray observations~\cite{auger} (see Fig.~\ref{fig:totalcrosssection}).
 
\section{Minimum Bias, Underlying Event, and Color Reconnection}
Minimum bias (MB) at 13 TeV is measured by ATLAS 
~\cite{atlas_mb}. 
The data is compared with predictions from \PYTHIA~(with A2~\cite{ATLAS:2011krm} and Monash~\cite{Skands:2014pea} tunes), \EPOS~and \QGSJET~\cite{Ostapchenko:2010vb}. \EPOS~provides the best description of the data, also as a function of $\sqrt{s}$ (in $|\eta|<0.2$). 
\PYTHIA~provides reasonable description of the data, while \QGSJET~performs the worst. 

Underlying event (UE) measurements in bins of several different variables are made by ATLAS at $\sqrt{s}=13$ TeV~\cite{ATLAS:2017blj}.
The measurements compared to predictions from \PYTHIA~(with A2, A14~\cite{TheATLAScollaboration:2014rfk}, and Monash tunes), \HERWIG~(UE-MMHT tune)~\cite{Bellm:2015jjp}, 
and \EPOS~show that the data can be described to $\sim$5\% accuracy, significantly larger than the data uncertainties of $\sim$1\%. Deviations of central values of the predictions in various observables indicate that improved MC tunes are needed for the interpretation of the LHC data. 
There is no best model, but in particular \EPOS, specialised for the simulation of inclusive soft QCD processes, displays significant discrepancies as the \pt~scale increases and thus may not be adequate for modelling multiple-parton interactions (MPI) at LHC. 

Sets of new CMS~\cite{ref:CMS} \PYTHIA~(CPi, i=1-5) and \HERWIG~(CHi, i=1-3) tunes~\cite{CMS:2019csb,CMS:2020dqt} are obtained assuming different \alpS$(M_Z)$ values used in the modelling of the initial-state and final-state radiation, hard scattering, and MPI, as well as the order of its evolution as a function of the four-momentum squared $Q^2$. The new tunes are distinguished according to the order of the NNPDF3.1 PDF set~\cite{Ball:2017nwa} used: LO, NLO, or NNLO. CP1 and CP2 are based on the LO, CP3 on NLO, and CP4 and CP5 on NNLO PDF set for \PYTHIA. All CHi tunes are based on NNLO-PDF with the corresponding \alpS$(M_Z)$ value for the PS component. The MPI and remnants components of CH1 uses NNLO-PDF with  \alpS$(M_Z)$=0.118; CH2 uses LO-PDF but with \alpS$(M_Z)$=0.118; and CH3 uses LO-PDF and its corresponding \alpS$(M_Z)$ of 0.130. 
Predictions of (N)NLO-PDF-based tunes reliably describe the central values of MB and UE data similar to LO-PDF tunes simultaneously with charged-particle distributions in diffractive and inelastic collisions. New tunes describe the data significantly better than the old tunes derived from data at lower collision energies. Double parton scattering is described worse by (N)NLO CPi tunes and none of the tunes describes the very forward region (-6.6$<$$\eta$$<$-5.2) better than $\sim$10\%. 
New tunes are tested also against \ttbar, \ttbar~jet shapes, 
Drell-Yan, dijet, V+jets for both CPi and CHi tunes, and also inclusive jet and event shape observables from LEP for the CHi tunes. 

UE activity in \ttbar~dilepton events are measured by CMS at $\sqrt{s}=13$ TeV
removing charged particles associated with the decay products of the \ttbar~event candidates as well as with pileup interactions 
for each event.
The observables and categories chosen for the measurements enhance the sensitivity to \ttbar~modelling, MPI, color reconnection (CR) and  \alpS($M_Z$) in \PYTHIA. 
Most of the comparisons indicate a fair agreement between the data and the 
\POWHEG~\cite{powheg}+\PYTHIA~setup with the \CP~tune, 
but disfavor the setups in which MPI and CR are switched off or
the default configurations of \POWHEG+\textsc{herwig++/7}, and \SHERPA~\cite{sherpa}.
The UE measurements in \ttbar~events test the hypothesis of universality of UE at an energy scale two times the top quark mass (\mt), considerably higher than the ones at which UE models have been studied in detail.
The results also show that a value of \alpFSR$(M_{Z})$=0.120$\pm$0.006
is consistent with the data and the corresponding uncertainties
translate to a variation of the renormalization scale by a factor of $\sqrt{2}$.

\begin{figure}
    \centering
    \includegraphics[width=8.0cm]{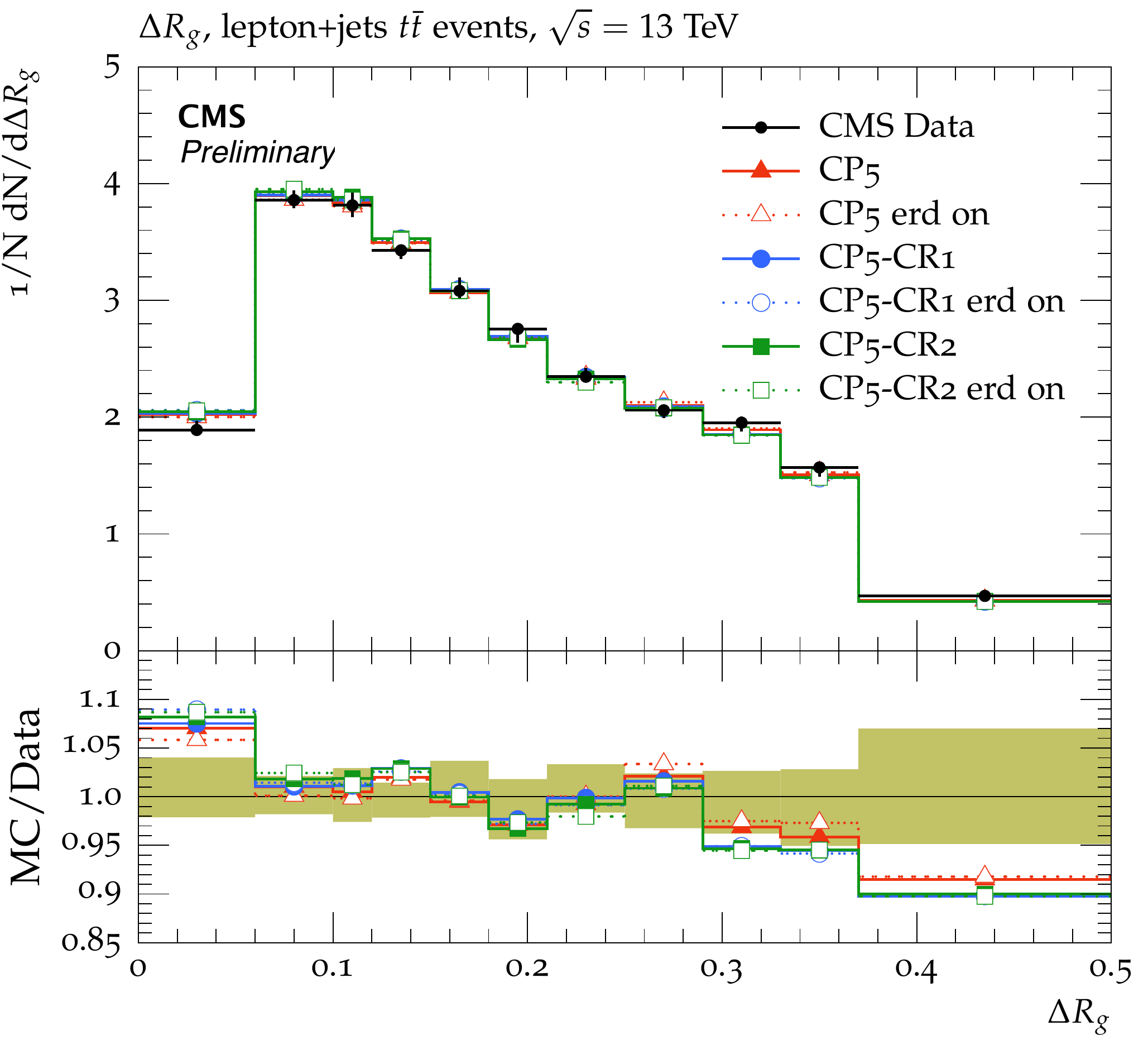}
    \caption{The normalized jet substructure angle between the groomed subjects. 
    Data are compared with the predictions from 
    CP5 and the corresponding CR tunes~\cite{CMS-PAS-GEN-17-002}. The colored band in the ratio plot represents the total experimental uncertainty in the data. 
    }
    \label{deltaRg}
\end{figure}

New sets of tunes for two of the CR models implemented in \PYTHIA, QCD-inspired (CR1) and gluon-move (CR2), are derived by ATLAS~\cite{ATLAS:2017wln} and more recently by CMS~\cite{CMS-PAS-GEN-17-002}. 
The ATLAS tunes are derived using ATLAS data taken at $\sqrt{s}=0.9$, 7, and 13 TeV, while CMS used $\sqrt{s}=$1.96 TeV CDF, and 7 and 13 TeV CMS data. 
It is observed that the new ATLAS CR models describe most observables within $\sim20\%$. 
The MPI-based CR and CR1 tunes of ATLAS perform significantly better than CR2. New CMS CR tunes are based on CP5 tune and they are tested against LEP, CDF, and 7-13 TeV data with MB, UE, forward energy flow, strange particle production, $p/\pi$, 
\ttbar~jet shapes (see Fig.~\ref{deltaRg}) and color flow in \ttbar. The new CMS CR tunes for MB and UE describe the data significantly better than the ones with the default parameters (also in the forward region). 
The new CMS CR tunes are also used to estimate the CR uncertainty on the 
\mt~measurement with semi-leptonic \ttbar~events. 
The largest deviation between the \mt~predictions is found to be between the CP5 and CP5-CR2 tune (with ERD) with a value of 0.32 GeV, similar to the value obtained in the dedicated \mt~measurement at 13 TeV (using CUETP8M2T4 tune~\cite{CMS:2016kle} and respective CR tunes). 
Therefore, CP5 with its new CR tunes does not improve or degrade the precision of the \mt~measurements and more detailed studies are needed. 
The CMS CR tunes are also tested against strange particle production. Although they describe well the rapidity distribution of $K_s^0$, they do not improve the description of the $\Lambda$ baryon rapidity distribution. 
This may indicate that improved hadronization models are needed. 
It is observed that baryon/meson ratios are well-described by the new CR tunes at different $\sqrt{s}$ of 10, 90 GeV, and 13 TeV. However, none of the tunes describe the $p/\pi$ yield ratio as a function of \pt~ (in the range \pt=0.4-1.2 GeV) in MB events at \text{13 TeV}. 
Moreover, none of the tunes describe the jet shapes well and with similar predictions. Some differences are observed with respect to color-flow data which is particularly sensitive to the ERD option in the CR models. It is verified that the impact of CR models is negligible in Drell-Yan events. To improve the description of Drell-Yan events higher-order corrections and improved MC codes are needed as shown by ATLAS~\cite{ATLAS-CONF-2021-033}; upgraded generators \MG(v.2.6.5) with FxFx merging simulating up to 3 jets at NLO, and \SHERPA(v.2.2.11) simulating up to 2 partons at NLO and up to 5 partons at LO and including NLO virtual electroweak corrections provide the best modelling.

\section{Top Quark Pair Spin Correlation and Common \ttbar~Setting for ATLAS and CMS}
The LHC Top Physics Working Group (LHC$top$WG) made comparisons of the ATLAS and CMS normalized cross sections in bins of $|\Delta\phi(\ell^+,\ell^-)|$ at the parton level\footnote{\url{https://twiki.cern.ch/twiki/bin/view/LHCPhysics/LHCTopWGSummaryPlots}}.
Very good agreement between ATLAS and CMS data and between ATLAS and CMS main MC predictions is observed. A good agreement of data with \MG~with FxFx merging~\cite{mg5,fxfx} is also observed as well as a fair agreement with the NNLO calculation \cite{behring19}. These comparisons pave the way for the first $\sqrt{s}$=13 TeV ATLAS+CMS combination from the LHC$top$WG.  In a related note ATLAS and CMS, presented the first studies towards a common \ttbar~MC settings~\cite{commonttbarmc_cmsatlas} that would facilitate ATLAS+CMS combinations and understanding the different MC configurations used in the two experiments.


\begin{thebibliography}{99}
\bibitem{amaldi73} U. Amaldi, \textit{et al.}, 
\href{https://doi.org/10.1016/0370-2693(73)90315-8}{Phys. Lett. B \textbf{44} (1973) 112.} 

\bibitem{froissart61}
M. Froissart, \href{https://doi-org/10.1103/PhysRev.123.1053}{Phys. Rev. 123 (1961) 1053.}

\bibitem{ref:ATLAS}
ATLAS Collaboration, \href{https://doi.org/10.1088/1748-0221/3/08/S08003}{JINST 3 (2008) S08003} 

\bibitem{ATLAS:2016ygv}
ATLAS Collaboration,
\href{https://doi.org/10.1103/PhysRevLett.117.182002}{Phys. Rev. Lett. \textbf{117} (2016) 182002.}

\bibitem{pythia8} T. Sjöstrand, \textit{et al.}, 
\href{https://doi.org/10.1016/j.cpc.2015.01.024}{Comput. Phys. Commun. {\bf 191} (2015) 159}.

\bibitem{Pierog:2013ria}
T.~Pierog, \textit{et al.}, 
\href{https://doi.org/10.1103/PhysRevC.92.034906}{Phys. Rev. C \textbf{92} (2015) 034906.}
\bibitem{auger} Pierre Auger Collaboration, 
\href{https://doi.org/10.1103/PhysRevLett.109.062002}{Phys. Rev. Lett. \textbf{109} (2012) 062002.} 
\bibitem{atlas_mb} ATLAS Collaboration, 
\href{https://doi.org/10.1016/j.physletb.2016.04.050}{Phys. Lett. B \textbf{758} (2016) 67.}
\bibitem{ATLAS:2011krm}
ATLAS Collaboration,
\href{https://cds.cern.ch/record/1400677}{ATL-PHYS-PUB-2011-014.}
\bibitem{Skands:2014pea}
P.~Skands, \textit{et al.}
\href{https://doi.org/10.1140/epjc/s10052-014-3024-y}{Eur. Phys. J. C \textbf{74} (2014) 3024.}
\bibitem{Ostapchenko:2010vb} S.~Ostapchenko,
\href{https://doi.org/10.1103/PhysRevD.83.014018}{Phys. Rev. D \textbf{83} (2011) 014018.}
\bibitem{ATLAS:2017blj}
ATLAS Collaboration,
\href{https://doi.org/10.1007/JHEP03(2017)157}{JHEP \textbf{03} (2017) 157.}
\bibitem{TheATLAScollaboration:2014rfk}
ATLAS Collaboration,
\href{https://cds.cern.ch/record/1966419}{ATL-PHYS-PUB-2014-021.}
\bibitem{Bellm:2015jjp}
J.~Bellm, \textit{et al.}
\href{https://doi.org/10.1140/epjc/s10052-016-4018-8}{Eur. Phys. J. C \textbf{76} (2016) 196.}

\bibitem{ref:CMS} CMS Collaboration, \href{https://doi.org/10.1088/1748-0221/3/08/S08004}{JINST 3 (2008) S08004} 

\bibitem{CMS:2019csb}
CMS Collaboration,
\href{https://doi.org/10.1140/epjc/s10052-019-7499-4}{Eur. Phys. J. C \textbf{80} (2020) 4.}
\bibitem{CMS:2020dqt}
CMS Collaboration,
\href{https://doi.org/10.1140/epjc/s10052-021-08949-5}{Eur. Phys. J. C \textbf{81} (2021) 312.}
\bibitem{Ball:2017nwa} R. D. Ball, \textit{et al.}, 
\href{https://doi.org/10.1140/epjc/s10052-017-5199-5}{Eur. Phys. J. C \textbf{77} (2017) 603.}
\bibitem{powheg} S. Alioli, \textit{et al.}, 
\href{https://doi.org/10.1007/JHEP06(2010)043}{JHEP {\bf 06} (2010) 43}. 
\bibitem{sherpa} E.~Bothmann, \textit{et al.},
\href{https://doi.org/10.21468/SciPostPhys.7.3.034}{SciPost Phys. \textbf{7} (2019) 034.}
\bibitem{ATLAS:2017wln}
ATLAS Collaboration, 
\href{https://cds.cern.ch/record/2262253}{ATL-PHYS-PUB-2017-008.}
\bibitem{CMS-PAS-GEN-17-002}
CMS Collaboration, 
\href{https://cds.cern.ch/record/2780467}{CMS-PAS-GEN-17-002.}
\bibitem{CMS:2016kle}
CMS Collaboration, 
\href{https://cds.cern.ch/record/2235192}{CMS-PAS-TOP-16-021.}
\bibitem{ATLAS-CONF-2021-033} 
ATLAS Collaboration,
\href{https://cds.cern.ch/record/2777239}{ATLAS-CONF-2021-033.}
\bibitem{mg5} J. Alwall, \textit{et al.}, 
\href{https://doi.org/10.1007/JHEP07(2014)079}{JHEP {\bf 07} (2014) 79}. 
\bibitem{fxfx} R. Frederix, S. Frixione, 
\href{https://doi.org/10.1007/JHEP12(2012)061}{JHEP {\bf 12} (2012) 61}. 
\bibitem{behring19} A. Behring, \textit{et al.}, 
\href{https://doi.org/10.1103/PhysRevLett.123.082001}{Phys. Rev. Lett. \textbf{123} (2019) 082001.}
\bibitem{commonttbarmc_cmsatlas}
ATLAS and CMS Collaborations,
\href{https://cds.cern.ch/record/2771088}{ATL-PHYS-PUB-2021-016}, \href{https://cds.cern.ch/record/2772793}{CMS-NOTE-2021-005.}
\end{thebibliography}
\end{document}